\documentclass[useAMS,usenatbib,mnras,letter]{mn2e}
\usepackage{aas_macros}
\usepackage{graphicx}
\usepackage{caption}
\usepackage{subcaption}
\usepackage{deluxetable}
\usepackage{amsmath}

\def\psr{PSR~J1717$-$4054}
\def\pasa{Publ. Astron. Soc. Australia}

\def\mujy{$\umu$Jy}
\def\nudot{$\dot{\nu}$}
\def\earth{\oplus}

\title[\psr{}]{The Three Discrete Nulling Timescales of \psr}

\author[M. Kerr, G. Hobbs, R. M. Shannon, M. Kicynski, R. Hollow, S.
Johnston]
{ M.~Kerr$^{1}$\thanks{E-mail: matthew.kerr@gmail.com},
  G.~Hobbs$^{1}$,
  R.~M.~Shannon$^{1}$,
  M.~Kiczynski$^{2}$,
  R.~Hollow$^{1}$,
  and S.~Johnston$^{1}$ \\
$^{1}$CSIRO Astronomy and Space Science, Australia Telescope National
Facility, PO~Box~76, Epping NSW~1710, Australia\\
$^{2}$School of Mathematical and Physical Sciences, University of Newcastle, University Drive, Callaghan NSW~2308, Australia}
\begin{document}

\date{Accepted 2014 August 14.  Received 2014 August 12; in original form 2014 June 16}

\pagerange{\pageref{firstpage}--\pageref{lastpage}} \pubyear{2014}

\maketitle

\label{firstpage}

\begin{abstract}
\psr{} is one of the small class of pulsars which null on intermediate
($\sim$hour) timescales.  Such pulsars represent an important link
between classical nullers---whose emission vanishes for a few
rotations---and the intermittent pulsars which null for months and
years.  Using the Parkes radio telescope and the Australia Telescope
Compact Array, we have studied the emission from \psr{} over intervals
from single pulses to years.  We have identified and characterised
nulling at three discrete timescales: the pulsar emits during ``active
states'' separated by nulls lasting many thousands of rotations, while
active states themselves are interrupted by nulls with a bimodal
distribution of durations---one to two rotations, or tens of
rotations.  We detect no subpulse structure or flux density variations
during active states, and we place stringent limits ($<0.1$\% of the
mean active flux density) on pulsed emission during inactive states.
Finally, our high-quality data have also allowed us to measure for the
first time many important properties of \psr{}, including its
position, spindown rate, spectrum, polarization characteristics, and
pulse broadening in the interstellar medium.

\end{abstract}


\section{Introduction}

Nulling pulsars, first identified by \citet{Backer70}, spontaneously
and suddenly (typically within one rotation) cease to emit at a
detectable level.  After an interval ranging from a few rotations
\citep[e.g.][]{Wang07} to a few years, e.g. PSR~J1841$-$0500
\citep[$\sim$600 days,][]{Camilo12}, emission resumes just as
suddenly.  This wide range of timescales, most of which are
significantly longer than those natural to a pulsar magnetosphere, is
difficult to explain through a single mechanism.

Because radio emission arises from coherent plasma processes, short
nulls have been interpreted as fluctuations in the plasma state that
briefly disrupt or alter the required coherency.  In other
pulsars---particularly those with drifting subpulses---the nulling
phenomenon may be related to the geometry of a drifting ``carousel''
of sparks \citep{Deshpande99,Rankin08} whose drift rate depends on
the local charge density above the polar cap.  For example,
\citet{Unwin78} determined the subpulse drift rate of PSR~B0809+74
drops to zero during nulls, while \citet{Herfindal07} identified
subpulse periodicity persisting through nulls in PSR~B1133$+$16,
suggesting these short nulls are simply gaps in the drifting beam
pattern.

On the other hand, the pulsars that null on long timescales, e.g.
PSR~B1931$+$24 \citep[$\sim$30 days,][]{Kramer06}, PSR~J1841$-$0500
\citep[$\sim$600 days,][]{Camilo12}, and PSR~J1832$+$0029
\citep[$\sim$600--800 days,][]{Lorimer12} are intermittent in
the sense that the currents and particle acceleration powering radio
emission may cease, or at the very least change sufficiently to steer
the radio beam away from the earth \citep[e.g.][]{Timokhin10}.  The
key piece of evidence comes from the measurement of the spindown rate
(\nudot{}) of each state, which decreases by a factor of 1.5--2.5
when the pulsar is nulling, indicating a dramatic reconfiguration of
the magnetospheric currents and, consequently, torque.

Similar switches between metastable magnetosphere/spindown states have
been identified by \citet{Lyne10}, who discovered a quasiperiodic
modulation of the spindown rate in a sample of pulsars monitored over
several decades.  Moreover, they found that in some pulsars, switches
between \nudot{} levels were accompanied by changes in pulse profile
\citep[i.e. mode changes,][]{Backer70B}, thus linking emission
properties to magnetosphere state.  The stochastic switches between
\nudot{} states introduce red noise into timing residuals relative to
a spindown model with a single \nudot{}, suggesting that such state
switching may be of fundamental importance to the timing noise
phenomenon and thus to pulsar timing arrays.

These long timescale observations make abundantly clear that at least
some pulsars switch between metastable magnetospheres.  However,
because switches happen rarely and rapidly, it is difficult to catch
these pulsars ``in the act'' and gain insight into both what these
states physically represent and how the switching is accomplished.
Pulsars with intermittency of hours likely also represent
magnetospheric switching, though the intermittent durations are too
short to measure independent values of \nudot{}.  Such pulsars can be
targeted with observations to resolve all relevant timescales and may
thus yield clues to the switching process.  In a clear example of the
power of such systems, \citet{Hermsen13} discovered that the X-ray
emission from PSR B0943$+$10 is correlated with radio mode changes,
implying rapid cooling of the neutron star polar cap after the pulsar
switches from its radio-faint to its radio-bright mode.

\psr{}, discovered by \citet{Johnston92} in a 20\,cm Parkes survey of
the Galactic plane, is a strong nuller and is inactive most of the
time.  Indeed, in a 2-hr observation, \citet{Wang07}
observed only one brief, 3.5 minute burst of emission, leading
those authors to conclude the nulling fraction is $>$95\%.  During
these scarce active states, the pulsar continues to null, albeit at a
much lower rate.  Hereafter, to distinguish between the multi-hour
nulls separating active periods (APs) and the $<$1-minute nulls within
active periods (see below), we refer to long nulls as inactive periods
(IPs).

With the lengthy breaks between APs, the telescope time required to
obtain many realizations of this cycle is prohibitive.  Instead, in
order to monitor the long-term behaviour of \psr{}, we used short, low
time-resolution snapshot observations obtained over six years with the
Parkes radio telescope.  To access short timescales, we complemented
these with two rise-to-set Parkes observations of 9.5\,hr during which
we obtained high time-resolution data.  During the first track, we
also observed \psr{} simultaneously with the Australia Telescope
Compact Array (ATCA).  With these data, we have characterised the
intermittency timescale and identified nulling at two discrete
timescales within APs; we have also measured for the first time many
important properties of \psr{}.

In \S\ref{sec:long}, we describe our monitoring observations and
discuss the resulting long-term timing solution, study of flux
stability, and measurement of typical AP/IP duration.  In
\S\ref{sec:short},  we describe our long-track Parkes and ATCA
observations, and we present in detail the emission properties of the
pulsar within its active periods.  We additionally use these
well-calibrated data to measure the polarization and pulse broadening
of \psr{} and to place a stringent limit on any persistent emission
from IPs.  We present a likelihood-based method for identifying pulse
nulls and present the resulting analysis of nulling within APs.
Finally, we summarise and interpret our results in
\S\ref{sec:discussion}.

\section{Monitoring Observations and Long-timescale Properties}
\label{sec:long}

\begin{deluxetable}{ll}
\tablewidth{\linewidth}
\tablecaption{\label{tab:parms} Measured and Derived Parameters for \psr{}}
\tablecolumns{2}
\tablehead{
\colhead{Parameter} &
\colhead{Value}
}
\startdata
Right ascension\tablenotemark{a}, R.A. (J2000.0)\dotfill&$17\mathrm{^h}17\mathrm{^m}$52\fs22(7) \\
Declination\tablenotemark{a}, decl. (J2000)\dotfill& --41\degr 03\arcmin17\arcsec(4)\\
Position epoch (MJD)\dotfill & 56671\\
Frequency, $\nu$ (Hz)\dotfill& 1.1264838005(3) \\
Frequency derivative, \nudot{} ($10^{-15}$\,Hz/s)\dotfill & --4.661(1)\\
Frequency epoch (MJD)\dotfill& 53200\\
Characteristic age, $\tau_c$\tablenotemark{b}\tablenotemark{c} (Myr)\dotfill& 3.8\\
Spindown luminosity,
$\dot{\mathrm{E}}$\tablenotemark{b}\tablenotemark{d} (erg\,s$^{-1}$)\dotfill&$2.1\times10^{32}$\\
Dipole magnetic field, B\tablenotemark{e} ($10^{12}$\,G)\dotfill&1.8\\
DM Distance (NE2001, kpc)\dotfill&4.7\\\tableline
Flux density, 732\,MHz (mJy)\dotfill& 17(1)\\
Flux density, 1369\,MHz (mJy)\dotfill& 5.2(2)\\
Flux density, 3094\,MHz (mJy)\dotfill& 1.0(1)\\
Flux density, 1400\, MHz, S$_{1400}$\tablenotemark{b} (mJy) \dotfill& 4.9\\
Spectral index, $\alpha$\dotfill& --1.9(1)\\\tableline
Dispersion measure, DM (pc\,cm$^{-3}$) \dotfill&306.9(1)\\
Rotation measure, RM (rad\,m$^{-2}$)\dotfill&--800(100)\\
Scattering timescale, $\tau_{sc,732}$  (ms)\dotfill& 60(6)\\
Scattering timescale, $\tau_{sc,1369}$ (ms)\dotfill& 7(1)\\
Scattering timescale $\tau_{sc,1000}$\tablenotemark{b} (ms)\dotfill&
20\\\tableline
Inactive fraction (\%)\dotfill& 80(15)\\
Active nulling fraction (\%) \dotfill& 6.8(3)\\
Mean active state (AP) duration (s)\dotfill&1100(180)\\
Mean inactive state (IP) duration (s)\dotfill&4300(800)\\
\enddata
\tablenotetext{a} {Position from ATCA imaging.}
\tablenotetext{b} {Derived.}
\tablenotetext{c} {$\tau_c\equiv-\nu/2\,\dot{\nu}$}
\tablenotetext{d} {$\dot{\mathrm{E}}\equiv-10^{45}\,\mathrm{erg\,s}^{-1}\,\nu\,\dot{\nu}/(2\pi)^2$}
\tablenotetext{e} {$\mathrm{B}\equiv-3.2\times10^{19}$\,G\,$\dot{\nu}/\nu^3$}
\tablecomments{Numbers in parentheses give the uncertainty on the
terminal significant figure(s).  The formal uncertainties on flux
densities and timescales are smaller than systematic uncertainties,
which we estimate to be about 10\%.}
\end{deluxetable}

\subsection{Observations}

The PULSE@Parkes program \citep{Hollow08} provides high school
students the opportunity to remotely control and observe pulsars with
the 64m Parkes radio telescope, and the data obtained are used for both
educational and scientific purposes \citep{Hobbs09}. 
During two-hour observing sessions, students work in 
small groups and select suitable pulsars from the program catalogue,
observing each pulsar for a few minutes.  As the project has
been running since December 2007, the data set provides an excellent
set of snapshot observations for a number of pulsars.  Because its
intermittent nature engages student interest, \psr{} is typically
observed in any session it is visible, resulting in 85 observations as
of April 2014.

PULSE@Parkes observations are largely undertaken with the 20\,cm
multibeam receiver \citep{Staveley-Smith96}, with 30\,s
sub-integrations of 256\,MHz of bandwidth centred about 1369\,MHz
recorded in 1024 frequency channels and 1024 phase bins by the Parkes
Digital Filterbank Mark 3 (PDFB3) or PDFB4, a nearly identical system.
Each pulsar observation is preceded by observation of a cycled,
coupled noise diode, allowing good polarization calibration via
measurement of differential gain and phase.

We primarily processed the data using the PSRCHIVE software suite
\citep{Hotan04}.  We mitigated aliased signals and narrowband
RFI by, respectively, excising channels within 5\% of the band
edge and those with a level substantially above a
median-smoothed bandpass. 
To convert the observed intensity to absolute flux density,
we made use of observations of the radio galaxy 3C 218 \citep[Hydra A;
][]{Baars77} taken every few
weeks to support the Parkes Pulsar Timing Array project
\citep{Manchester13} to determine the system equivalent flux density
of the calibration noise source and consequently an
absolute flux scale for each Stokes parameter.

All data recorded through the PULSE@Parkes program become public
immediately and are available through the CSIRO Data Archive
Portal\footnote{https://data.csiro.au} \citep{Hobbs11}.  Data taken
under other programs, including those described below, are generally
available after an 18-month proprietary period.

\subsection{Timing Solution}

\begin{figure}
\includegraphics*[angle=270,width=0.45\textwidth,viewport=50 -30 550 670]{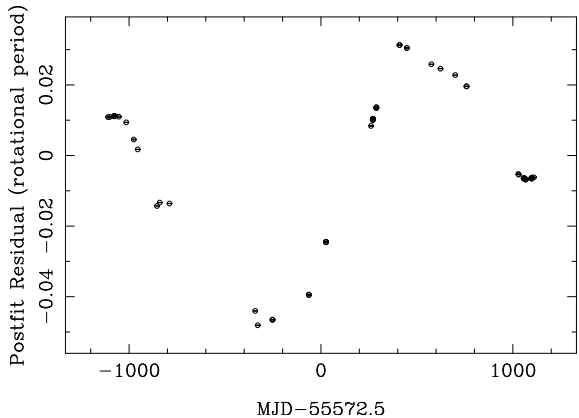}
\caption{\label{fig:timing_solution}Residuals of 20\,cm pulse times of
arrival in fractions of a neutron star rotation for the best-fit
timing model.}
\end{figure}

Combining these data with archival observations, we
used the \textsc{Tempo2} software package \citep{Hobbs06} to obtain a
phase-connected timing solution extending over six years.  The residuals
to a simple fit for the frequency ($\nu$) and its derivative
(\nudot{}) show modest timing noise and appear in Figure
\ref{fig:timing_solution}.  This is the first published measurement of
\nudot{}, from which several important properties, listed in Table
\ref{tab:parms}, can be estimated.

\subsection{Pulsar Profile and Flux Density Variations}

To measure the active-state flux density, we selected only
observations in which emission was detected, and we further deleted
sub-integrations visibly affected by substantial nulling or a state
switch.  We employed two methods of calculating
the flux density: (1) simply adding up signal in excess of baseline
and (2) fitting an analytic template to the profile and inferring the
flux from the template scale.

\begin{figure}
\includegraphics[scale=0.42]{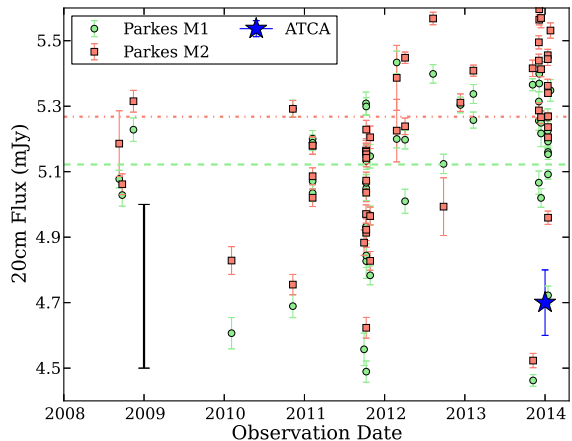}
\caption{\label{fig:flux_lc}The absolute flux density of \psr{} at
20\,cm.  The green circles and dashed line (mean) indicate flux
densities derived using Method 1, while the salmon squares and
dot-dashed line (mean) indicate flux densities derived with Method 2,
i.e. by fitting a template.  An uncertainty of 10\%---a typical level
encapsulating calibration errors and interstellar scintillation---is
indicated by the black error flag in the lower lefthand corner.  The
blue star indicates the ATCA-derived pulsed flux density
(\S\ref{sec:ATCA}).}
\end{figure}

The results, shown in Figure \ref{fig:flux_lc}, indicate the intrinsic
flux density is stable over the years of observation, with variations
readily explained by modest scintillation in the interstellar medium
and small calibration errors.  Moreover, the flux density is mututally
consistent between the two methods; since the second measurement
explicitly measures flux density with the expected pulse shape while
the first method measures any flux density above baseline, the pulse
profile must also be stable at this level.

The mean flux density is 5.2$\pm$0.3\,mJy, which is substantially
different from previous values reported in the literature.
\citet{Johnston92} found 1.0\,mJy from the discovery observation, but
this computation did not select only APs, and is consistent with
scaling our value by the inactive fraction.  \citet{Hobbs04b} give a
flux density of 54\,mJy based on data from the Parkes multibeam pulsar
survey.  Because the survey pointings were substantially offset from
the position of \psr{}, this discrepancy may have resulted from either
an inaccurate model of the sensitivity at the edges of the beam or the
use of an inaccurate pulsar position.

We also obtained calibrated flux measurements from the 10\,cm and
50\,cm observations discussed below, with flux densities of 0.96\,mJy
and 16.9\,mJy, respectively, following a fairly typical power law
$S_{\nu}=S_0(\nu/\nu_0)^{\alpha}$ with $\alpha=-1.9$.

\subsection{Active/Inactive Duty Cycle}
\label{sec:nf}

With many observations of \psr{}, we are in a position to refine the
nulling fraction (inactive fraction, in our parlance) estimate of
\citet{Wang07}.  However, the typical time \psr{} spends in either
state is substantially longer than the few-minute PULSE@Parkes
observations.  Further, as there is a natural bias among both high
schools students and astronomers to extend (curtail) observations of
the pulsar when it is active (inactive), a simple tally of the total
time the pulsar is observed to be bright is biased.

On the other hand, switches are random, and, if the process is
memoryless (Poisson), switch rates are unaffected by observational
bias.  Moreover, since switches are rare for this pulsar, we expect to
see no more than one state switch in a given sub-integration.  Thus,
to estimate switch rates from our fold-mode data, which has typical
time resolution of 30\,s, we classify by inspection each
sub-integration into one of four classes: $IP\rightarrow AP$,
$IP\rightarrow IP$, $AP\rightarrow AP$, or $AP\rightarrow IP$.  We
then simply count the instances of each class and tally the total time
spent in each state in Table \ref{tab:switches}.  Averaged over all
PULSE@Parkes and archival observations, we find a rate of
$IP\rightarrow AP$ of $(2.3\pm0.4)\times10^{-4}$\,Hz and
$AP\rightarrow IP$ of $(9.3\pm1.6)\times10^{-4}$\,Hz.  That is, the
typical time the pulsar spends in the active (inactive) state is
$1100\pm180$\,s ($4300\pm800$\,s), and the inactive fraction is
$80\pm15$\%.  Although limited by the small number of observed
switches, we note that these rates appear to be consistent from year
to year.  We also note that the nulling fraction obtained by simply
tallying time in each state is 77\%, lower than, but consistent with,
the estimate from switch counting.

\begin{deluxetable}{lrrrrr}
\tablewidth{\linewidth}
\tablecaption{\label{tab:switches}State switches from active to
inactive and
vice versa observed, as well as total time recorded in each state, in PULSE@Parkes and archival data.}
\tablecolumns{6}
\tablehead{
\colhead{Year} &
\colhead{IP$\rightarrow$AP} &
\colhead{AP$\rightarrow$IP} &
\colhead{IP} &
\colhead{AP} &
\colhead{AP Frac}\\
\colhead{} &
\colhead{} &
\colhead{} &
\colhead{(s)} &
\colhead{(s)} &
\colhead{(\%)}
}
\startdata
2004 & 8 & 9 & 52584 & 14626 & 22 \\
2005 & 3 & 3 & 15948 & 3904 & 20 \\
2006 & 0 & 1 & 3682 & 1374 & 27 \\
2007 & 2 & 5 & 14095 & 3757 & 21 \\
2008 & 4 & 3 & 7326 & 3494 & 32 \\
2009 & 0 & 0 & 2544 & 0 & 0 \\
2010 & 2 & 2 & 2819 & 1193 & 30 \\
2011 & 5 & 10 & 8159 & 5383 & 40 \\
2012 & 3 & 2 & 15331 & 1994 & 11 \\
2013 & 3 & 1 & 6252 & 2944 & 32\\ \tableline
All & 30  & 36 & 128740 & 38658 & 23 \\ 
\enddata
\end{deluxetable}

\section{Single-pulse Observations and Short-timescale Properties}
\label{sec:short}

\begin{figure}
\centering
\begin{subfigure}[b]{0.45\textwidth}
    \includegraphics[angle=270,width=\textwidth]{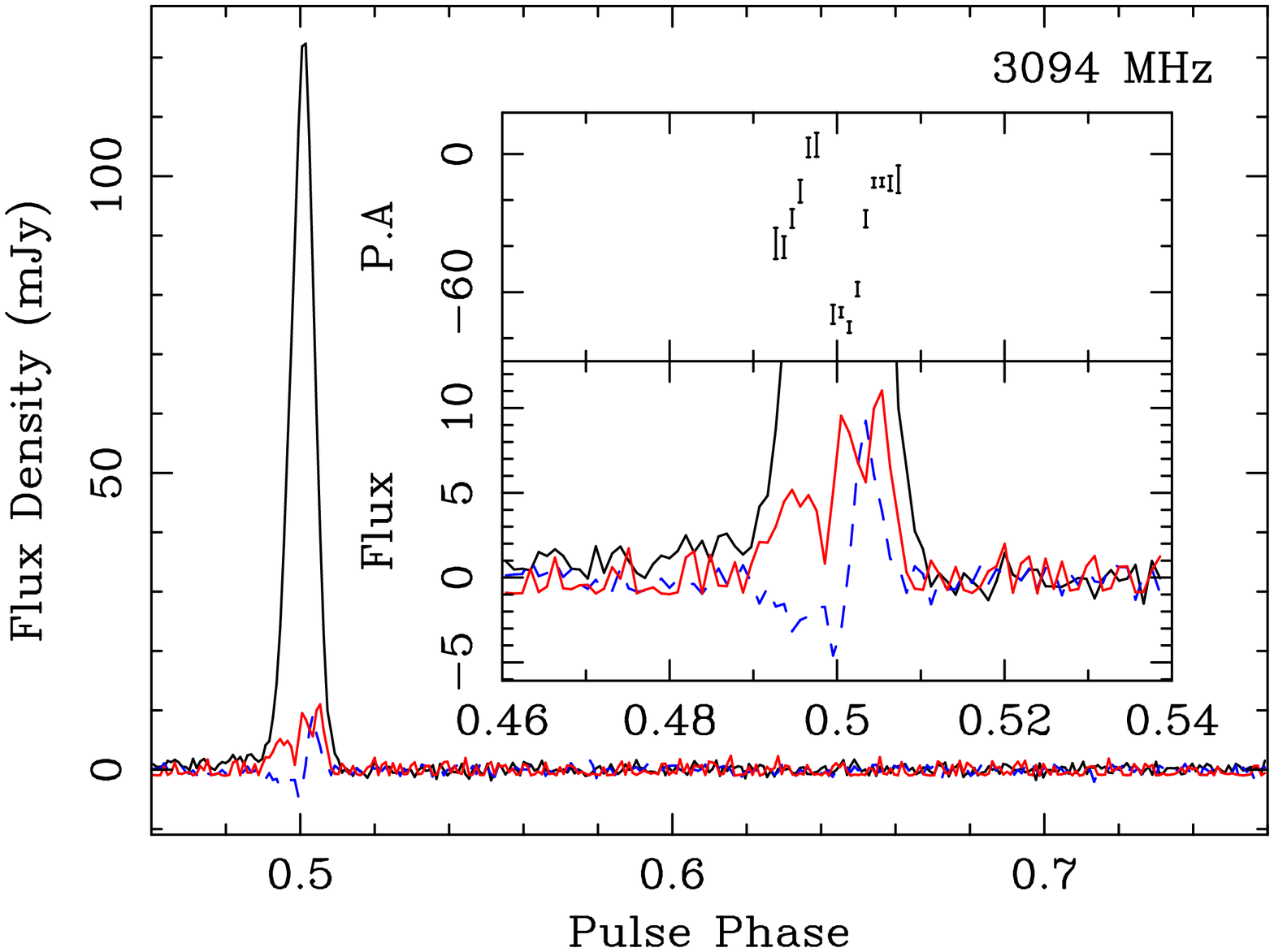}
\end{subfigure}\\
\begin{subfigure}[b]{0.45\textwidth}
    \includegraphics[angle=270,width=\textwidth]{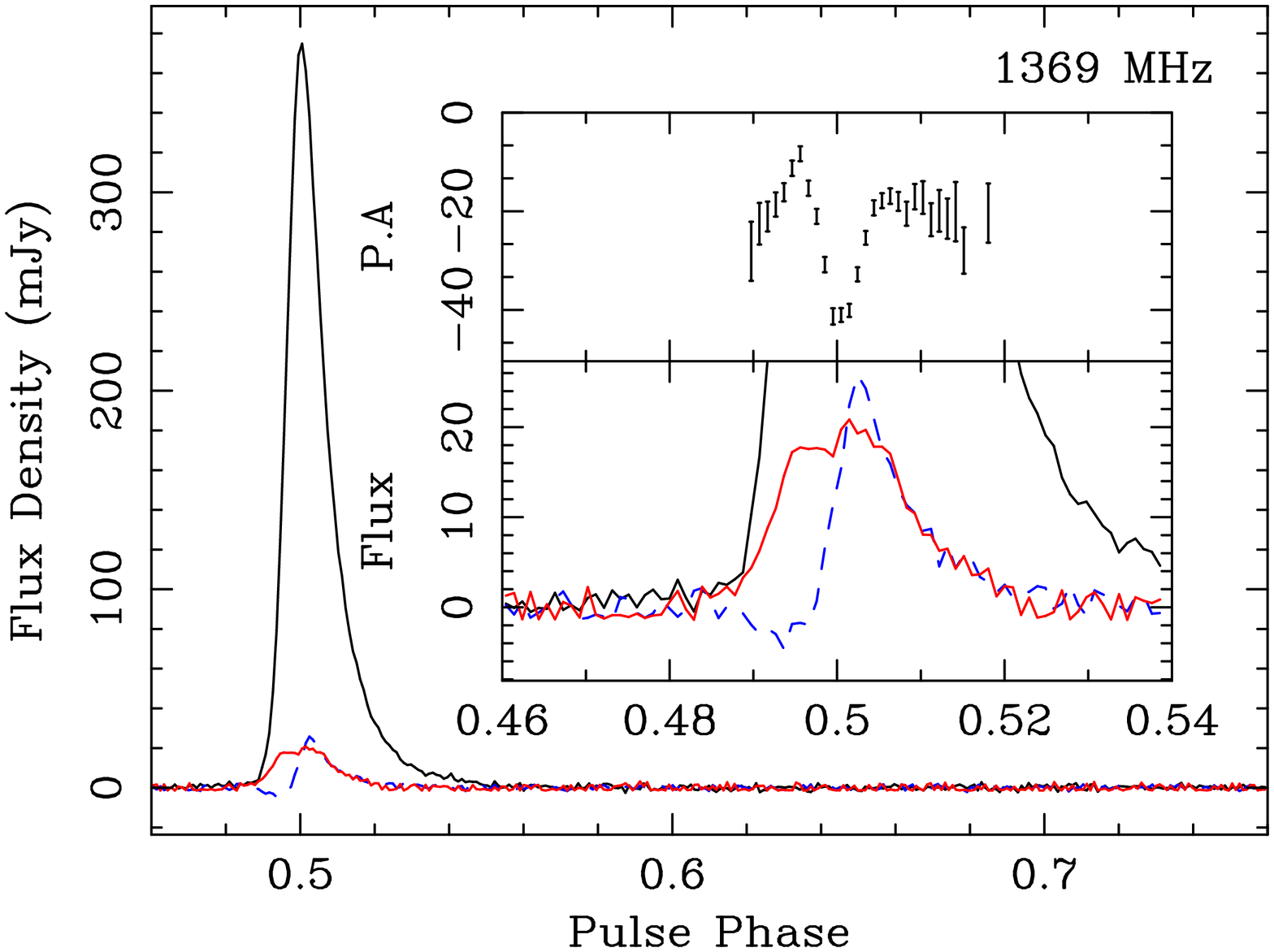}
\end{subfigure}\\
\begin{subfigure}[b]{0.45\textwidth}
    \includegraphics[angle=270,width=\textwidth]{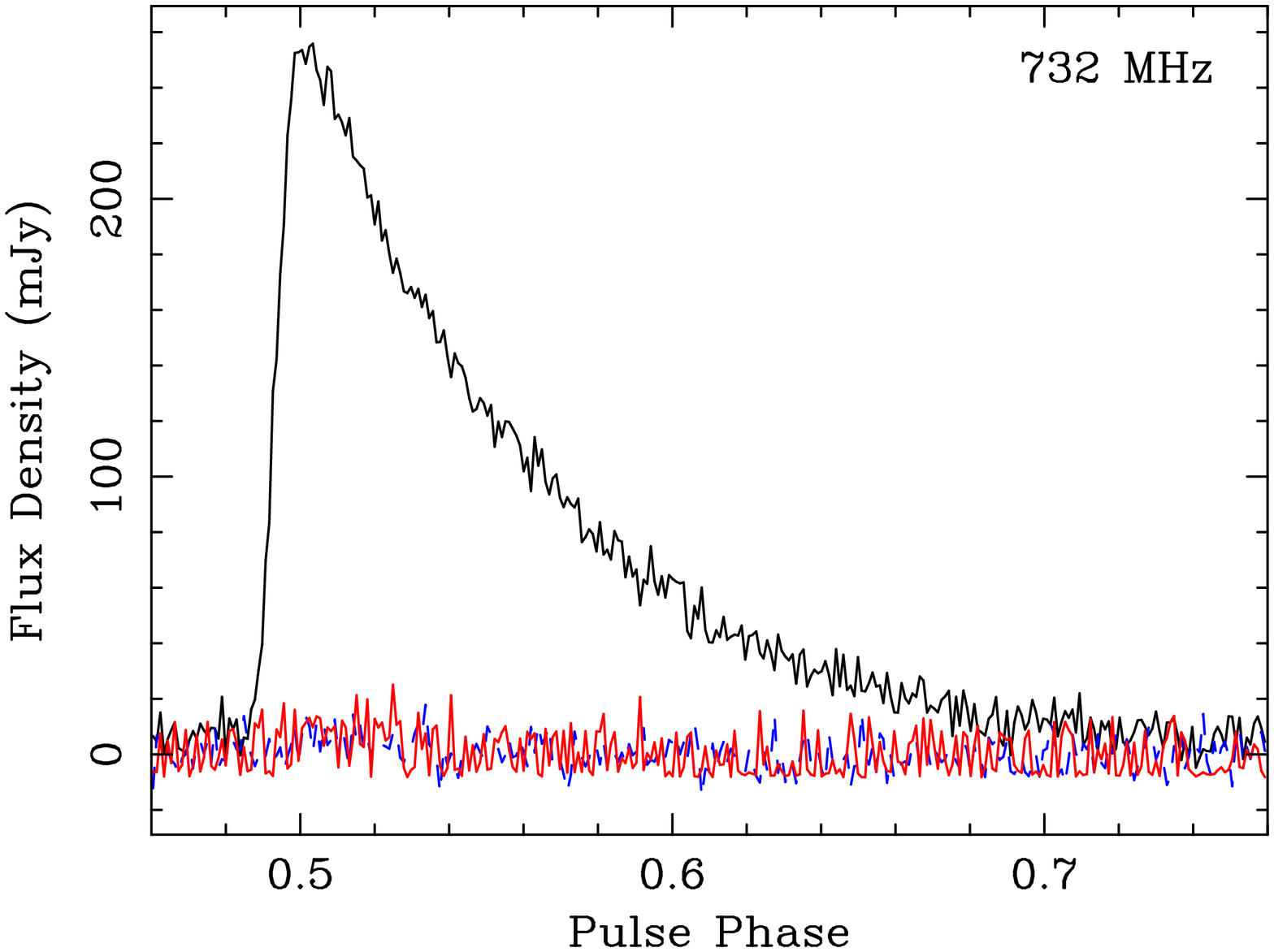}
\end{subfigure}%

\caption{\label{fig:profiles}Pulse profiles at (top to bottom) 10\,cm,
20\,cm, and 50\,cm.  Total intensity (Stokes I) is the dominant black
trace.  A small amount of circular polarization (Stokes V) appears in
dashed blue at 10\,cm and 20\,cm, as does a similar level of linear
polarization (solid red).  The inset in the top panel shows a zoomed
view of the profile, in which appears a hint of a leading component.}

\end{figure}

\begin{figure}
\includegraphics[width=0.45\textwidth]{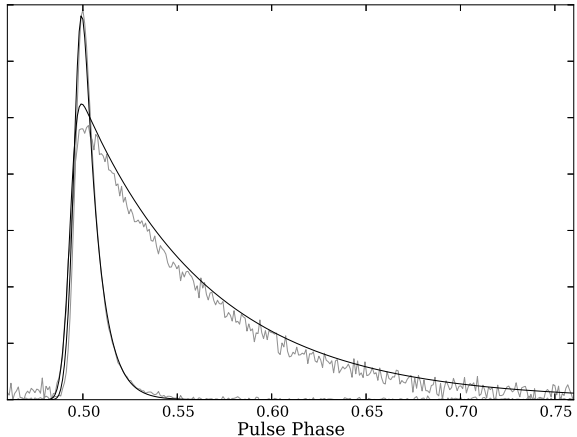}
\caption{\label{fig:scattering}Stokes I profiles at 20\,cm
(grey, sharp) and 50\,cm (grey, long tailed) along with models of the
scattered profile (black) as described in the text.  The scale is
arbitrary.}
\end{figure}

\subsection{Observations}

\subsubsection{Parkes}

To resolve state switches and record the entirety of a series of APs
and IPs, we obtained high-time-resolution observations (Parkes program
P850) of \psr{} during two 9.5\,hr rise-to-set tracks on 13 (Day 1)
and 14 (Day 2) January, 2014.  On Day 1, we observed at 20\,cm with
the same receiver configuration described above.  With PDFB4 in
``search mode'', we recorded a 2-bit, 256-channel filterbank every
256\,$\umu$s, while simultaneously we
obtained fold-mode data with PDFB3.  To maintain polarization
calibration, we activated the pulsed noise source roughly every hour. 


On Day 2, we collected data using an identical search-mode
configuration but with a dual-band 10\,cm/50\,cm receiver
\citep{Granet05}, offering 1024\,MHz of bandwidth at a centre
frequency of 3094\,MHz and 64\,MHz of bandwidth at 732\,MHz.  As the
two bands required both backends to take search-mode data, no
fold-mode data were obtained.  We reduced all search-mode data to 1024-bin single pulse profiles with \textbf{dspsr} \citep{vanStraten11}.

\subsubsection{ATCA}
\label{sec:ATCA}
Concurrent with the 20\,cm Parkes observations on Day 1, we also
observed \psr{} with the six-element ATCA interferometer for 5.5\,hr.
We recorded 512 channels of data over a total bandwidth of 2048\,MHz
centred at 2102\,MHz using the Compact Array Broadband Backend in
pulsar-binning mode \citep{Wilson11}, which produces visibilities in
32 phase bins each 10\,s correlator cycle.  Primary, bandpass, and flux calibration were achieved using the
radio galaxy PKS~1934$-$638, which has little
time-dependent flux variation.  Secondary phase calibration and gain
calibration were conducted with the radio galaxy PKS~1714$-$397, and we
used self-calibration to correct for phase variations
of the antennas during the observation.

The observations covered the final three APs of Day 1 (see Figure
\ref{fig:waterfalls}).  To measure the pulsar flux, we first
subtracted the off-pulse visibilties from the on-pulse visibilties,
removing the non-pulsed flux in the field.  We inverted these
visibilities to form a dirty image and subsequently cleaned the image
using a multi-frequency algorithm.

We measured the flux density across the entire band and also in
interference-free sub-bands, and we fit our broad-band observations to
derive a spectral index of $-2.1\pm0.08$, consistent with those
obtained with Parkes data.  We estimate the AP flux density to be
$4.7\pm0.1$~mJy at~1369~MHz, about 10\% lower than the Parkes
measurement.  Since the ATCA flux measurements include nulls within
the AP, while the Parkes measurements do not, we expect the former to
be lower by about 10\%.  Further, the Parkes and ATCA flux scales are
derived from two different sources (Hydra A and PKS~1934$-$638); and
\psr{} was about 500\arcsec{} from the phase centre of the ATCA
observations, making the flux measurement particularly dependent on
the primary beam model for the receiving system.

\subsection{Extended and Off-pulse Emission Limits}
\label{sec:off-pulse}
Using the ATCA images, we searched for but detected neither extended
nor pointlike emission in the off-pulse phase bins.  The former could
arise, e.g., from a pulsar wind nebula.  These typically
form around pulsars with large energy outputs (strong winds) and/or
those with high space velocities embedded in dense environments
\citep{Gaensler06}.  Although \psr{} sits in the Galactic plane
and its velocity is unknown, it has a relatively low spin-down
luminosity, and our nondetection is unsurprising.

Pointlike off-pulse emission might appear from faint magnetospheric
emission (e.g. from higher altitudes) or from reflection of pulsed
radiation from a disk of objects (asteroids) with size much larger
than radio wavelength \citep{Phillips93,Cordes08}.  \citet{Cordes08}
found the ratio of unpulsed to pulsed flux to be
\begin{equation}
\label{eq:asteroid}
\frac{S_u}{S_p} =  10^{-1.5} \left( \frac{A}{0.5} \right) \frac{f_g
M_{a,-4}}{\rho\,r_{10}^2 \tilde{R}_{a,2}},
\end{equation}
where $A$ is the material albedo, $f_g$ is a beaming function,
$M_{a,-4}=M_a/10^{-4} M_\earth $ is the mass of the asteroid belt,
$r_{10} = 10^{10} r$~cm is the radius of the asteroid belt, and $
\tilde{R}_{a,2}$ is the root mean square (r.m.s.) radius of the asteroids.  The beaming
factor $f_g$ is the unknown ratio of fluence beamed in the direction
of the asteroid belt to the ratio beamed in the direction of Earth.
By imaging the off-pulse bins both during APs and over the entire
observation, we place an AP limit of 120\,\mujy{}\,beam$^{-1}$
and 33\,\mujy{}\,beam$^{-1}$ overall.  The
former limit is about $10\%$ of the pulsed flux and does not constrain
the properties of any circumpulsar material.  On the other hand, if
the pulsar's intermitency is caused by a modest shift of the pulsar
beam away from the earth while the disk illumination remains
unchanged, the latter limit applies and begins to probe the predictions of Eq. \ref{eq:asteroid}.

\subsection{Pulse Profiles and Polarimetry}
To obtain high S/N pulse profiles, appearing in Figure
\ref{fig:profiles}, we co-added the APs throughout the long tracks,
using the fold-mode data available at 20\,cm and the folded
search-mode data for 10\,cm and 50\,cm.

At 10\,cm, the profile is a narrow, approximately gaussian pulse.
Linear polarization is present at low levels, while the modest
circular polarization appears to flip handedness in the pulse center.
The polarization position angle has a classic ``S'' swing, but the
pulse is too narrow to constrain the magnetic inclination.  There is
some evidence for a leading component of comparable width to the
dominant, visible peak.  The 20\,cm profile is similar to the 10\,cm
profile, save for the appearance of a scattering tail.  By dividing
the 20\,cm and 10\,cm bands into eight sub-bands and minimizing the
residuals to $\delta t\propto\mathrm{DM}\,\nu^{-2}$, we determined the
dispersion measure (DM) to be $306.9\pm0.1$\,pc cm$^{-3}$.  We note
that this is not necessarily the absolute DM, as the fit is covariant
with profile evolution and an unknown time offset between the two
bands.

The profile is highly scattered at 50\,cm, and any intrinsic
polarization signal is smeared away.  To measure the scattering
timescale, $\tau_{sc}$, we convolved the 10\,cm profile with an
exponential function $\exp(-t/\tau_{sc})$ and varied $\tau_{sc}$ until
we obtained a reasonable representation of the lower frequency data
(Figure \ref{fig:scattering}).  We find values of
$\tau_{sc,732}=60$\,ms and $\tau_{sc,1369}=7$\,ms at 732\,MHz and
1369\,MHz.  We estimate a systematic uncertainty due to approximating
the intrinsic profile with the 10\,cm profile of about 10\%.  If
$\tau_{sc}\propto\nu^{-\beta}$ describes the relation of scattering
time and frequency, we find $\beta=-3.4\pm0.4$, somewhat less than the
canonical value $\beta=-4$.  Extrapolating the 732\,MHz measurement to
1\,GHz, we obtain $\tau_{sc}=21$\,ms.  We note that the NE2001 model
of the electron distribution in the Galaxy \citep{Cordes02} predicts a
$\tau_{sc}$ of only 1.0\,ms at 1\,GHz, indicating substantial
scattering structure in excess of the model prediction along the line
of sight.


\subsection{Faint Pulsed Emission in Inactive State}

We examined the 20\,cm fold-mode data for the presence of pulsed
emission while the pulsar was in the inactive state.  After excising
APs and sub-integrations contaminated by impulsive broadband RFI, we
retained 6.6\,hr which we reduced to a 128-bin Stokes I
profile with an r.m.s. of $180\pm14$\,\mujy{}.   The excellent
agreement with the predicted radiometer-noise limit of 169\,\mujy{},
indicates the absence of both a pulsed signal and any substantial
RFI.  Using the mean 20\,cm profile as a template, we obtain a
2$\sigma$ upper limit on emission from the inactive state of
3.9\,\mujy{}, less than 0.1\% of the mean AP flux.

We likewise searched the ATCA images for on-pulse emission during IPs,
and we set an upper limit of 12\,\mujy{}, consistent with that above.
To detect emission with a different pulse profile, we also formed the
difference between the two halves of the pulse-phase bins, but we
again found no evidence for pulsed emission.


\begin{figure*}
\centering
\begin{subfigure}[b]{0.45\textwidth}
    \includegraphics[bb=23 70 526 526,clip]{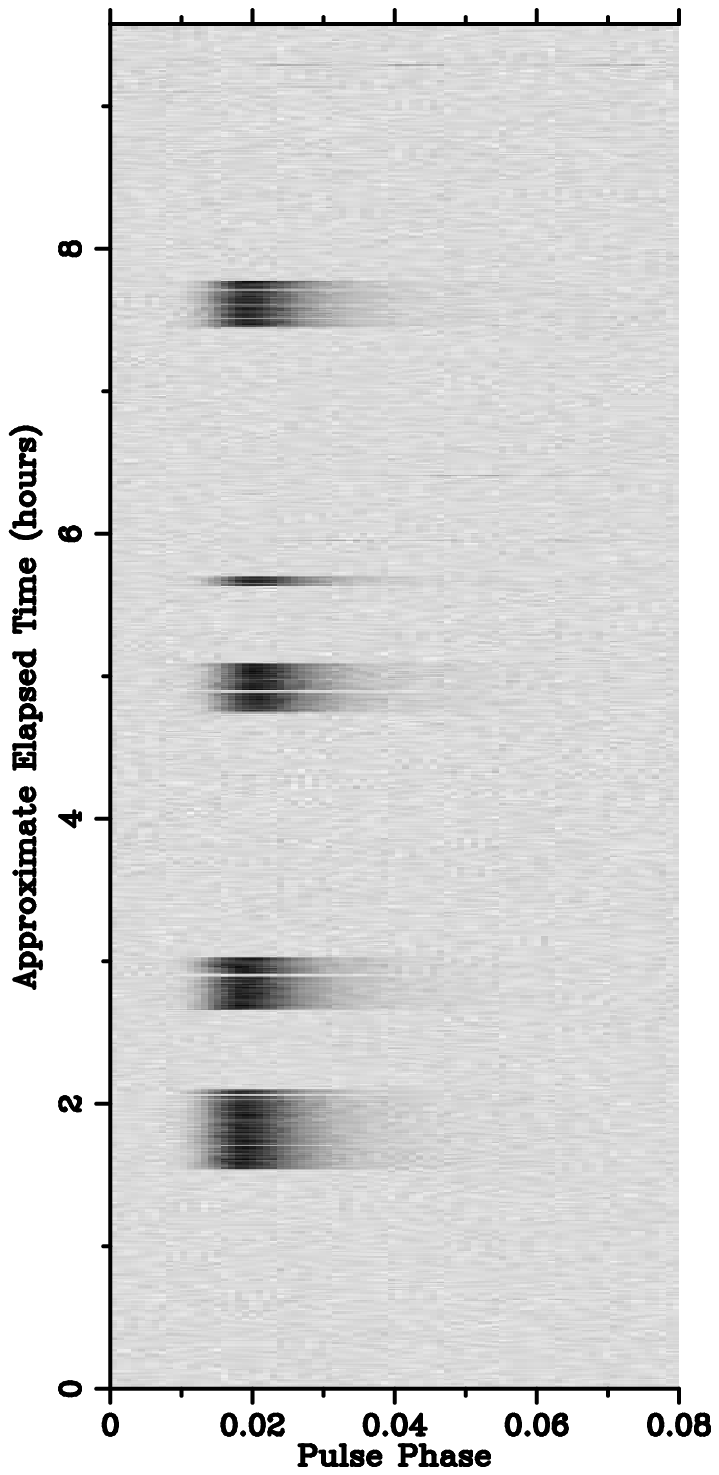}
    \caption{20\,cm / Day 1.}
\end{subfigure}%
\begin{subfigure}[b]{0.45\textwidth}
    \includegraphics[bb=23 70 526 526,clip]{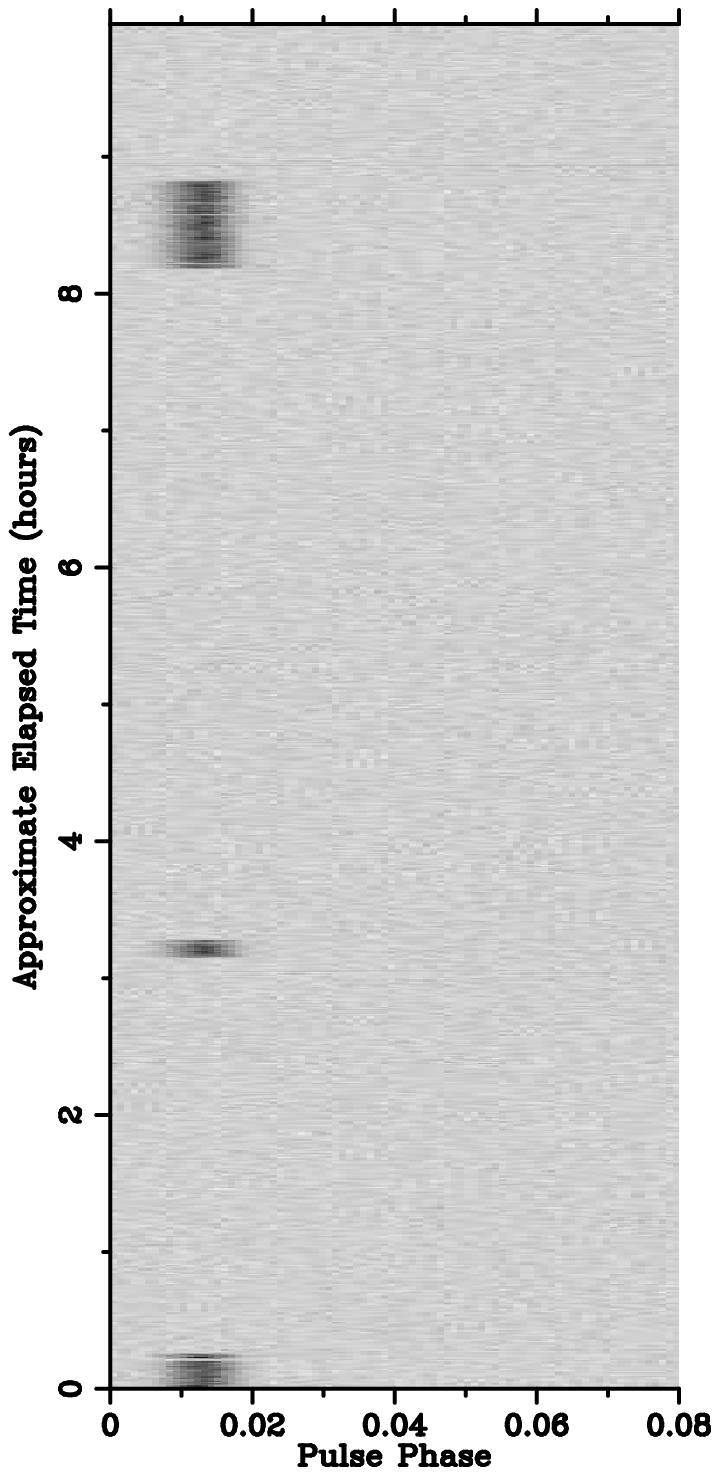}
    \caption{10\,cm / Day 2.}
\end{subfigure}%
\caption{\label{fig:waterfalls}The phase-resolved intensity (arbtirary
units) obtained during the two days of long-track
observations, demonstrating the stochastic nature of the intermittency.}
\end{figure*}

\begin{figure*}
\includegraphics[width=\textwidth]{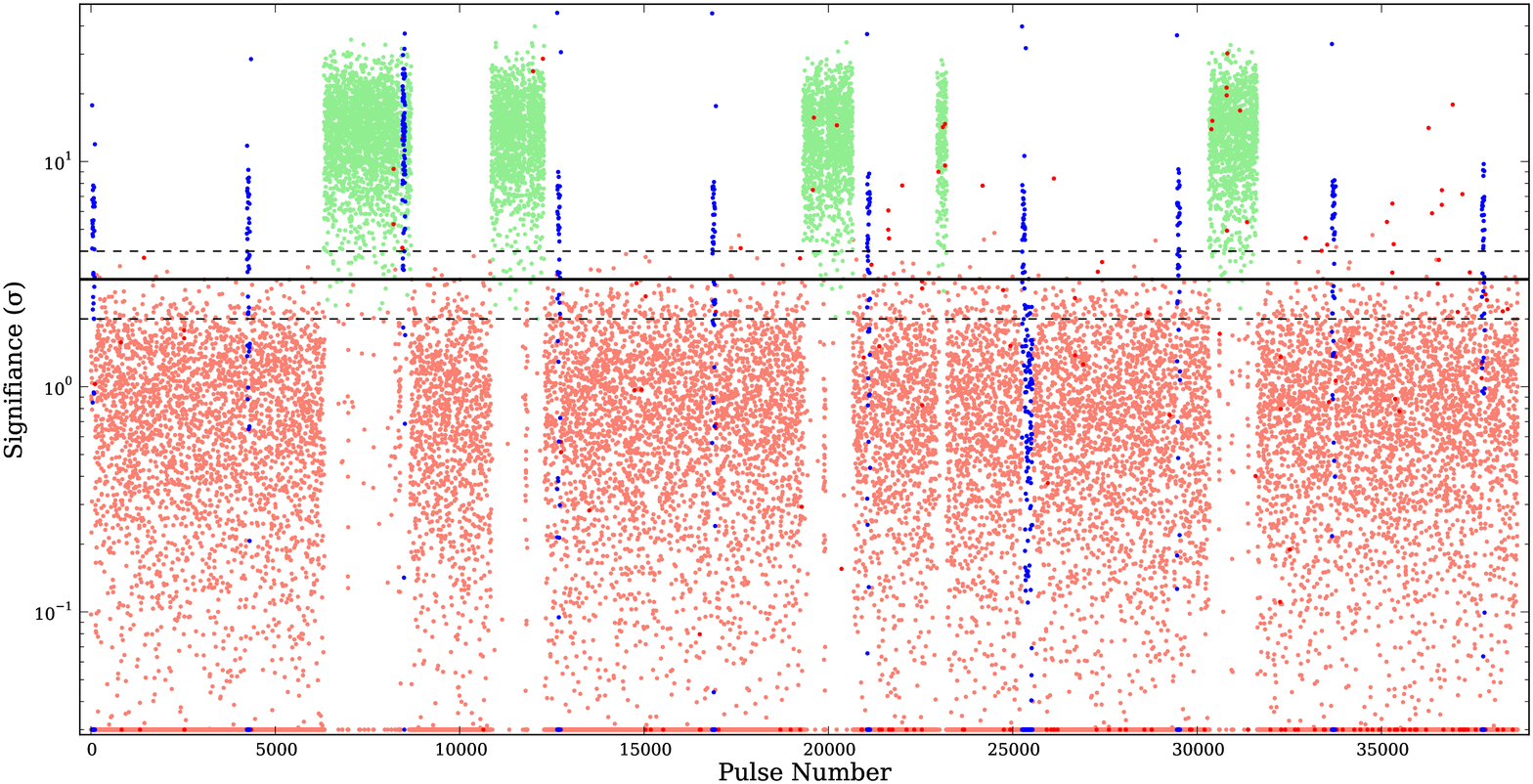}
\caption{\label{fig:ts}$\sqrt{\mathrm{TS}}$ for 20\,cm single pulse
profiles.  ``ON'' (green) and ``OFF'' (salmon) are classified
according to the algorithm described in the text.  Values with
$\sqrt{\mathrm{TS}}<0.1$ are set to 0.1\,$\sigma$ for display purposes
and appear at the bottom of the figure.  Pulses affected by
the presence of the calibrator are coloured in blue, while those
flagged as suffering from impulsive RFI are drawn in red.  The solid
black line is drawn at 3$\sigma$, the dashed lines at 2$\sigma$ and
4$\sigma$.  In rotations, the five active states
measure 2363, 1436, 1352, 263, and 1320, the intervening inactive
states 2172, 7019, 2285, and 7105, and the bookending inactive states
$>$6320 and $>$7059.}
\end{figure*}

\subsection{Time-resolved Active States}

\subsubsection{Active/Inactive Duty Cycle}
\label{sec:aid}

During the 20\,cm observations of Day 1, the pulsar switched on for
five discrete APs with durations given in Figure \ref{fig:ts}; see
also Figure \ref{fig:waterfalls}.  Including only complete active and
inactive intervals, the mean AP lasts 1347 rotations, or 1196\,s, while
the mean IP lasts 4645 rotations, or 4124\,s, consistent with our
estimates in \S\ref{sec:nf}, albeit with much poorer statistics.
Since the bounding nulls must be at least as long as the observed
values, including these intervals increases the mean null duration to
$>$4729\,s.

On Day 2 (10\,cm and 50\,cm observations), we observed only two full
APs (durations of 450\,s and 2220\,s; see Figure \ref{fig:waterfalls})
and one partial AP at the onset of the observation ($>$870\,s).  The
IPs within the observation were much longer (9960\,s and 16,980\,s)
than those of Day 1.  The IP/AP duty cycle observed during the long
track observations disfavour a memoryless activation process, as there
appear to be no short IPs.


\subsubsection{AP Nulling}

\begin{figure*}
\centering
\begin{subfigure}[b]{0.48\textwidth}
    \includegraphics[width=\textwidth]{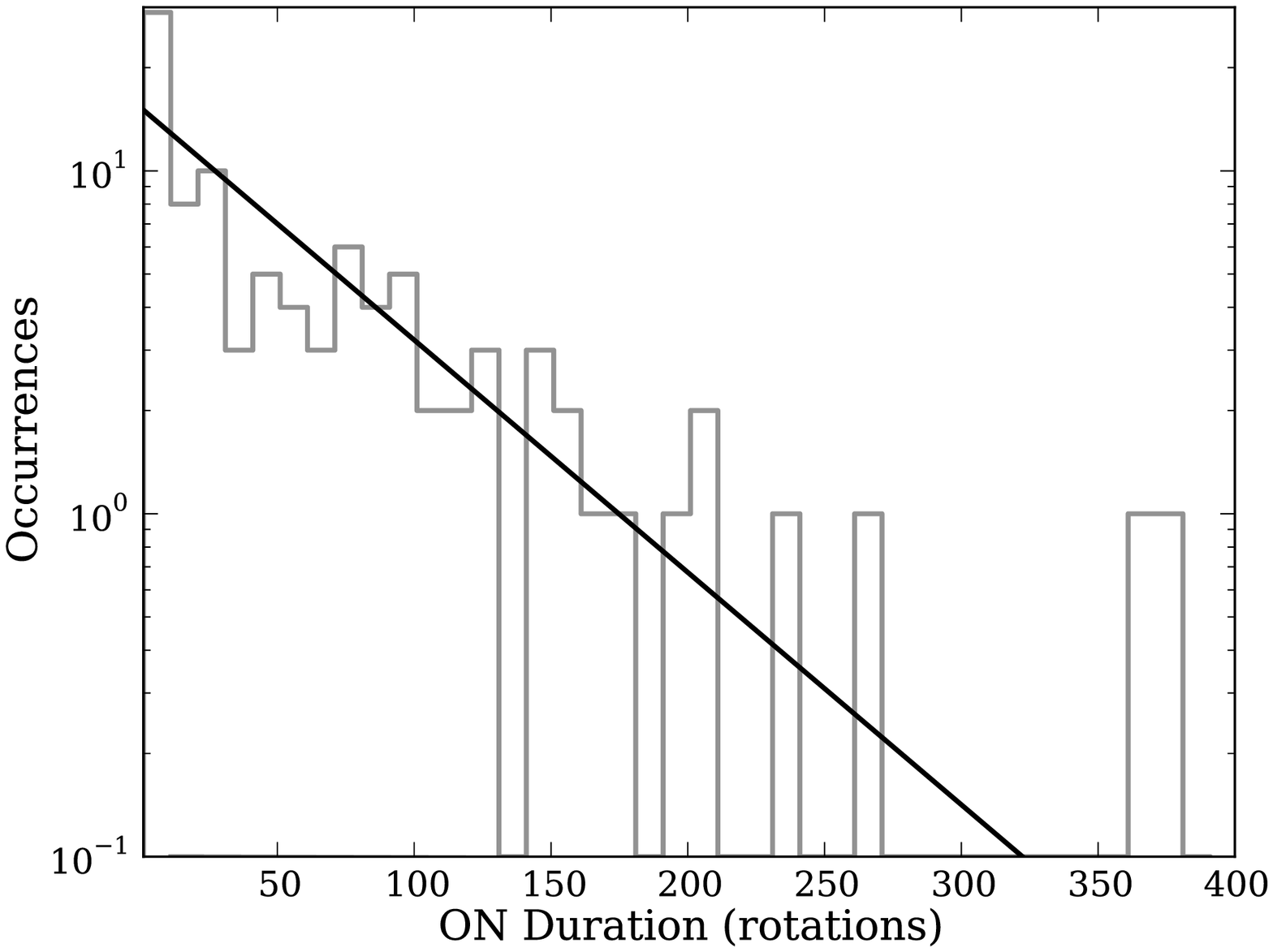}
\end{subfigure}%
\begin{subfigure}[b]{0.48\textwidth}
    \includegraphics[width=\textwidth]{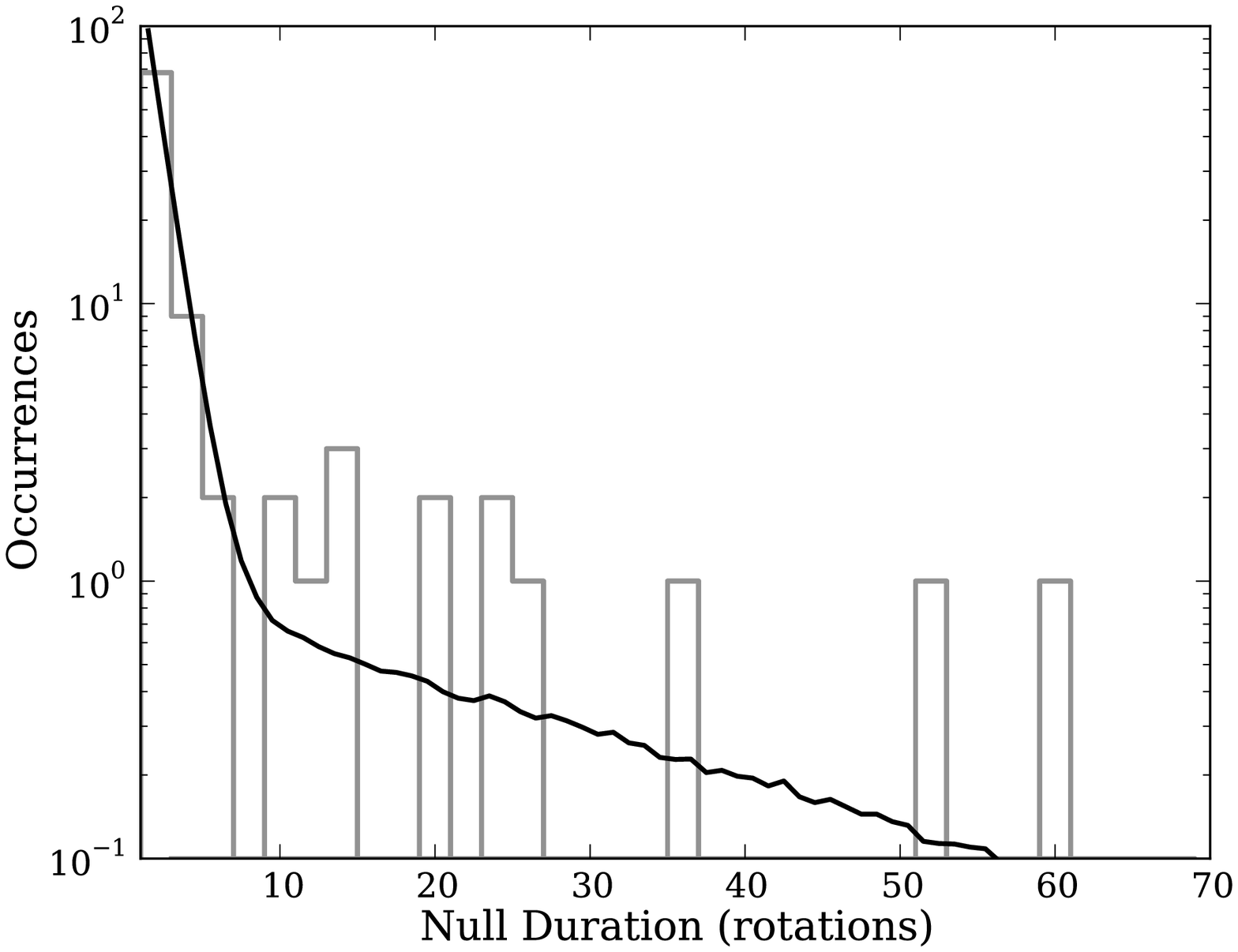}
\end{subfigure}%
\caption{\label{fig:intra}
Duration of non-nulling (grey, left) and nulling (grey, right)
intervals during active periods.  The non-nulling durations are shown
with the best-fit exponential distribution overplotted in black, while
the model shown in black in the right panel is the mean value from
simulations from a two-state Markov model described in \S\ref{sec:markov}.}
\end{figure*}

Within APs, \psr{} continues to null.  To identify null pulses, we
derived an analytic template from the high S/N 20\,cm profile
described above, and we fit this template to each single pulse.
Although we noted some variation in pulse shape, the mean profile was
generally an acceptable description of the single pulse emission.  We
estimated the baseline from the offpulse mean, and the phase was known
from the timing solution, leaving the signal strength, $s$, of the
template in each pulse as the only free parameter.  We adopted twice
the log likelihood ratio for $s=\hat{s}$, the best-fit value, versus
$s=0$ as our test statistic (TS), setting $\mathrm{TS}=0$ when
$\hat{s}<=0$.  In the absence of a signal, $\mathrm{TS}=0$ half of the
time, while the remaining $\mathrm{TS}>0$ values follow a $\chi^2$
distribution with one degree of freedom.  Then,
$\sqrt{\mathrm{TS}}=\sigma$, i.e.  the chance probability to observe
$\mathrm{TS}\geq\sigma$ is the tail probability of a standard normal
distribution integrated from $Z=\sigma$.

By examining data from the long null periods, we found the mean TS to
be about 10\% higher than expectations, but that rescaling by $0.9$
gave excellent agreement with the expected $\chi^2$ distribution.
This modified TS, expressed in $\sigma$ units, is calculated for each
single pulse and shown in Figure \ref{fig:ts}.  Note that we only
expect about 3 pulses in 1000 from the IP to exceed a TS of 3$\sigma$,
which is roughly the rate we observe.  To identify nulls within the
APs, we apply the following classification:
\begin{itemize}
\item any pulse with $\mathrm{TS}>4\sigma$ belongs to an AP, as
the chance of a null pulse exceeding this threshold is negligible;
\item any pulse with TS below $2\sigma$ is a null; although this
threshold is arbitrary, since the TS distribution in the alternative
hypothesis has an unknown distribution, we justify it below;
\item if a pulse has $2\sigma<\mathrm{TS}<3\sigma$, it is a null unless both adjacent pulses are from an AP;
\item if a pulse has $3\sigma<\mathrm{TS}<4\sigma$, it belongs to an
AP unless both adjacent pulses are nulls.
\end{itemize}

Using this classification, within the APs, we observed a total of 6281
pulses and 458 nulls, for an active period nulling fraction of
$6.8\pm0.3$\%.

\subsubsection{AP Null Duration Distribution}
\label{sec:markov}
The distribution of null and non-null durations within the APs is
shown in Figure \ref{fig:intra}.  The distribution of non-null
durations is approximately exponential, indicating that the pulsar
does not ``remember'' how long it has been shining since the previous
null.  The slight excess of short ``on'' states relative to the model
may stem from incorrect classification of a pulse as a null (though
see below).

The distribution of nulls, on the other hand, is poorly fit by an
exponential, and appears to be bimodal.  Short nulls of a few
rotations \citep[Type I as classified by][]{Backer70} significantly
outnumber the longer (Type II) nulls extending for tens of rotations.
Since the TS distributions of null and non-null pulses are not
perfectly separated, some of the Type I nulls may simply be faint
pulses.  To check this, we co-added the 52 single-pulse nulls and
found TS$=3.58\sigma$, indicating that a few of the nulls may be
pulses, but that the majority are \textit{bona fide} nulls.  A similar
analysis of 16 two-pulse nulls yields TS$=0$, while co-adding Type II
nulls with durations of $>5$ rotations yields TS$=3.1$, indicating the
long nulls may include a few faint pulses.

To further characterize the nulling process, we attempted to model it
as a three-state Markov process following \citet{Cordes13}.  In such a
process, the probability for the system to switch from one state to
another is described by a transition matrix $Q$.  The diagonal entries
$q_{ii}$ give the probability for the system to remain in its current
state and are related to the mean occupancy time of a state; the
off-diagonal elements are transition frequencies.  In general, such a
process produces monotonically decreasing distributions of state
occupancy times, tending to exponential distributions as
$q_{ii}\rightarrow1$.  Using the observed values for null duration and
frequency, we adopted a transition matrix
\begin{equation*}
Q =
 \begin{pmatrix}
  0.9844 & 0.0137 & 0.0019 \\
  0.5848 & 0.4142 & 0.0000 \\
  0.0386 & 0.0000 & 0.9614
 \end{pmatrix},
\end{equation*}
i.e. a single ``on state'' (state 0) and a short-lived (state 1) and a
long-lived (state 2) null state.  The observational
indistinguishability of states 1 and 2 is represented by a forbidden
transition $q_{12}=q_{21}=0$.  We simulated many realisations of the
process and show the resulting distribution of null durations in
Figure \ref{fig:intra}.  The three-state Markov model is largely
acceptable, though it fails to reproduce the gap observed between the
short-lived and long-lived nulls.  (This is a general property of such
models, as the probability density function for the duration of any
state monotonically decreases.)

\subsubsection{Single Pulse Flux Density Distribution}

\begin{figure}
\includegraphics[width=0.5\textwidth]{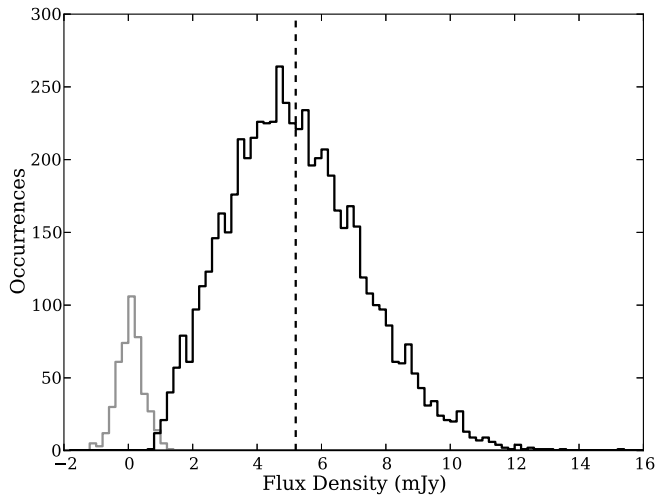}
\caption{\label{fig:fluxdist}Distribution of observed single pulse
flux densities; those identified as nulls are shown in grey, while
those identified as non-nulls are drawn in black.  The mean flux,
5.2\,mJy, is indicated by the dashed vertical line.}
\end{figure}

The value of $s$ we obtained to determine TS corresponds directly to
the flux density of each single pulse, modulo some scatter from
pulse-to-pulse profile variations.  The distribution of AP pulse flux
densities, excluding pulses affected by RFI and scaled to the measured
mean flux (5.2\,mJy), appears in Figure \ref{fig:fluxdist}.  The nulls
form a narrow normal distribution peaked about zero.  This
distribution appears identical to the flux density distribution
obtained from IPs (not shown), which is exceedingly gaussian and has
an r.m.s. of 0.39\,mJy.  The flux density distribution from non-null
pulses, with its high-flux tail, appears to be approximately
lognormal, though we note there is a slight underrepresentation of
faint pulses relative to that model.  Because the variance of the IP
distribution is small compared to the non-null distribution (r.m.s.
2.05\,mJy), the observed distribution is a good proxy for the
intrinsic distribution of single pulse fluxes.

\subsubsection{Pulse-to-pulse Correlations}

Finally, we searched the data for evidence of correlations between
pulses, e.g. as from drifting subpulses.  As the relatively modest S/N
and the narrow pulse window preclude direct detection of subpulses, we
instead searched directly for periodicity in the power spectral
density (PSD) of the single pulse fluxes.  However, the PSD of each of
the APs was consistent with white noise, and we conclude there is no
measurable correlation between pulses.  Likewise, we looked for any
correlation between null duration and the flux of the
preceeding/following pulses, e.g. as in that of PSR~B1944$+$17
\citep[e.g. ][]{Deich86}.  There is no evidence of correlation, save
perhaps that the first pulse of an AP tends to be weaker than average.

\section{Summary and Discussion}
\label{sec:discussion}

In summary, we find \psr{} is well described as a four-state system:
an ``active'' state (0) with nulls at two discrete timescales (1 and
2) and no evidence of complex pulse substructure or memory across
nulls; and an ``inactive state'' (3) whose emission, if present, must
be at least 1,000 times fainter than that of state 0. It remains
unclear if any of the null states (1, 2, and 3) correspond to the
same magnetospheric configuration.  The stability of the pulsar's flux
density and profile over timescales of years, as well as the absence
of pecularities in its timing solution, suggest the pulsar is stable
when averaged over many state transitions, i.e. more than a few days.

The distribution of durations of both APs and IPs has a large scatter,
but due to the lack of short APs and (especially) short IPs, each
appears to have an intrinsic timescale---that is, neither switching
process is memoryless.  The complete APs and IPs in our two long-track
observations have mean values of 1240\,s and 7240\,s, respectively,
and standard deviations of 690\,s and 5160\,s, indicating a fairly
white spectrum for the IPs and a redder spectrum for the APs.  Both
timescales are difficult to reconcile with any typical magnetospheric
process, and a more likely scenario is a switch between two metastable
states triggered by a quasiperiodic perturbation.  For example,
\citet{Cordes08} have proposed that circumpulsar debris from supernova
fallback disks might perturb a pulsar's magnetosphere, particularly
the outer gaps \citep{Cheng86} or the equatorial disk of return
current \citep[e.g.][]{Spitkovsky06}.  Such a disk could lie well
below our limits on reflection (\S \ref{sec:off-pulse}) but still
source asteroids which plunge into the magnetosphere every few hours
and provide enough ionized material to substantially alter the current
flow.  Although the ionized material would only persist for about a
rotation, the magnetosphere would at that point have settled into a
new metastable state.  In this picture, short AP nulls are simply due
to plasma fluctuations or patchy emission, an interpretation supported
by the approximately exponential distribution of ``on'' intervals.
The longer ($\sim$10 rotation) nulls, however, likely represent a
state switch, and may represent an intermediate state between active
and inactive states: if the long nulls and IPs correspond
to identical magnetospheric configurations, the absence of nulls of of
intermediate lengths (few hundred rotations) is puzzling.

Although we cannot directly measure the difference in spindown rate
between APs and IPs, we can estimate the contribution of such
switching to the r.m.s. of the TOA residuals to our timing solution
(Figure \ref{fig:timing_solution}).  Following Equation 12 of
\citet{Cordes13} and using the measured state durations of 1100\,s
(active) and 4300\,s (inactive) and the corresponding state
probabilities 0.2 and 0.8, the estimated contribution
$\sigma\approx1.7\,\mathrm{ms}\,\delta\dot{\nu}/\dot{\nu}$.  For the
observed r.m.s. $\approx16$\,ms, this implies
$\delta\dot{\nu}/\dot{\nu}\approx9$, somewhat larger than values
measured from intermittent pulsars (1.5--2.5), implying other sources
of timing noise are also important, though \citet{Timokhin10} suggests
that relative spindown rates may depend sensitively on small changes
in magnetosphere geometry.

With the measurement of \nudot{} we can now place \psr{} in the
context of other nulling pulsars.  Despite having one of the highest
known NFs, the spindown luminosity ($2\times10^{32}$\,erg\,s$^{-1}$)
and characteristic age (3.8\,Myr) of \psr{} \textbf{are entirely
unremarkable}.  C.f., e.g., Table 1 of \citet{Wang07}.  On the other
hand, \citet{Biggs92}, in an analysis of 43 nulling pulsars, found a
strong (but scattered) correlation between NF and pulse period, and
\citet{Wang07} observe a modest correlation with age.  Thus, the high
NF of \psr{} is somewhat anomalous.  As noted by \citet{Cordes08},
large NFs seems to occur at shallow magnetic inclinations, implying
\psr{} may have a magnetic inclination $<$45\degr{}.  Excluding the
very long nulling (intermittent) pulsars such as PSR~B1931$+$24,
\psr{} also has one of the longest inactive periods.  If, on the other
hand, one only considers nulling within active periods, the NF drops
to a much more modest 6.8\%, in line with many of the other nullers.

It is useful to compare \psr{} directly with a few other high NF
pulsars with long nulls.  In particular, recent work by
\citet{Gajjar14} details the properties of PSRs~J1738$-$2330 and
J1752$+$2359, both with NFs near 90\%.  Strikingly, both pulsars
exhibit features that seem to be absent in our similarly sensitive
observations: correlated burst onsets, a decline in flux over time
within a burst, and evidence of emission during the long nulls.  In
contrast, \psr{} seems simply to switch on and off.  Likewise,
PSR~B1944$+$17 \citep{Kloumann10} sports a high (nearly 70\%) NF with
nulls up to $\sim$100 periods.  Its null frequency peaks at one
rotation, however, suggesting the short nulls may be due to a carousel
pattern.  This may be the case for \psr{}, though we have no
additional evidence through subpulse structure.  Finally, it is also
worth pointing out PSR~J1502--5653, a scaled Doppelg\"{a}nger of
\psr{}; its $\mathrm{\dot{E}}$ and $\tau_c$ agree to within a factor
of two, and its NF is similarly high, 93\% \citep{Wang07}.  It, too,
displays a pattern of IP/nulling AP, save with IP and AP durations
scaled down by a factor of 10, making it a tempting target for
long-track study.

In conclusion, we now have a detailed picture of \psr{} on all
timescales of interest and have also measured a panoply of important
properties.  Substantial advance---e.g., identifying pulse
substructure---must likely await the the large collecting area of the
SKA, though some interim progress might be made with, e.g., ultrawide
bandwidth feeds, or with additional long-track observations.  The
identification of nulling at three timescales (few pulses, tens of
pulses, and thousands of pulses) is challenging to interpret in any
single picture, and we hope these observations will stimulate the
introduction of new physical models for state switching.


\section*{Acknowledgments}

We thank the anonymous referee and the journal editors, whose effort
and input improved this paper.

The Parkes radio telescope and Australia Telescope Compact Array are
part of the Australia Telescope, which is funded by the Commonwealth
Government for operation as a National Facility managed by CSIRO.

PULSE@Parkes is funded by CSIRO Astronomy and Space Science.  We are
grateful to the Australia-Japan Foundation, the Victorian Space
Science Education Centre, SPICE at the University of Western
Australia, the Astrophysics groups at the University of Cardiff and
University of Oxford, ASTRON, NAOJ Mizusawa VLBI Observatory, Koriyama
Space Park, Penrith Anglican College, and the University of
Brownsville, Texas for support in funding and hosting sessions. The
authors would like to acknowledge the more than 1,100 students from 93
schools who have taken part in sessions and contributed to the data
gathering since the program's inception in December 2007, as well as
the CSIRO Astronomy and Space Science staff, visitors, and
co-supervised PhD students whose volunteer participation has enriched
the program.

\label{lastpage}

\end{document}